\documentclass{amsart}  
\usepackage{amsmath,amssymb,latexsym,hyperref,url,graphicx,verbatim}

\newtheorem{definition}{Definition}

\graphicspath{{figures/}}

\newcommand{\colonequal}{\mathrel{\mathop:}=}

\newcommand{\wc}{\square}
\newcommand{\bc}{\blacksquare}
\newcommand{\s}{\mspace{-2.76mu}}
\newcommand{\icontriple}[3]{\normalsize\rule{0ex}{2ex}$\mspace{-4mu}#1\s#2\s#3\mspace{-4mu}$}
\newcommand{\iconsinglec}[1]{\normalsize$#1$}

\newcommand{\singleblackcell}{\cdots\wc\s\wc\s\wc\s\bc\s\wc\s\wc\s\wc\cdots}
\newcommand{\singlewhitecell}{\cdots\bc\s\bc\s\bc\s\wc\s\bc\s\bc\s\bc\cdots}
\newcommand{\twoblackcells}{\cdots\wc\s\wc\s\wc\s\bc\s\bc\s\wc\s\wc\s\wc\cdots}

\begin{document}

\title[Boundary growth in one-dimensional cellular automata]{Boundary growth in \\ one-dimensional cellular automata}

\author{Charles D. Brummitt}
\address{Department of Mathematics \& Complexity Sciences Center\\
	University of California\\
	One Shields Avenue\\
	Davis, CA 95616, USA}

\author{Eric Rowland}
\address{LaCIM, Universit\'e du Qu\'ebec \`a Montr\'eal, Montr\'eal, QC H2X 3Y7, Canada}

\date{\today}

\begin{abstract}
We systematically study the boundaries of one-dimensional, 2-color cellular automata depending on $4$ cells, begun from simple initial conditions. We determine the exact growth rates of the boundaries that appear to be reducible. Morphic words characterize the reducible boundaries. For boundaries that appear to be irreducible, we apply curve-fitting techniques to compute an empirical growth exponent and (in the case of linear growth) a growth rate. We find that the random walk statistics of irreducible boundaries exhibit surprising regularities and suggest that a threshold separates two classes. Finally, we construct a cellular automaton whose growth exponent does not exist, showing that a strict classification by exponent is not possible.
\end{abstract}

\maketitle

\section{Introduction}
Cellular automata are simple machines consisting of cells that update in parallel at discrete time steps. In general, the state of a cell depends on the state of its local neighborhood at the previous time step. The earliest known examples were engineered for specific purposes, such as the two-dimensional cellular automaton constructed by von Neumann in 1951 to model biological self-replication~\cite{von Neumann}. 
Three decades later, researchers began to study entire classes of automata, such as the $256$ one-dimensional cellular automata that use $k=2$ colors and that depend on $d=3$ cells~\cite{Wolfram 1983, Wolfram 1984}. The behavior of these rules subsequently garnered much attention. Most studies have focused on the interiors of patterns generated by cellular automata, likely because the boundaries are well known and simple for the $k=2$, $d=3$ rules, such as the three linear boundaries shown in Figure~\ref{rules_90,30,110}. However, boundaries of automata are diverse, often more predictable than interiors (and hence more amenable to mathematical study), and even useful for detecting interesting behavior.

\begin{figure}
	\scalebox{.3}{\includegraphics{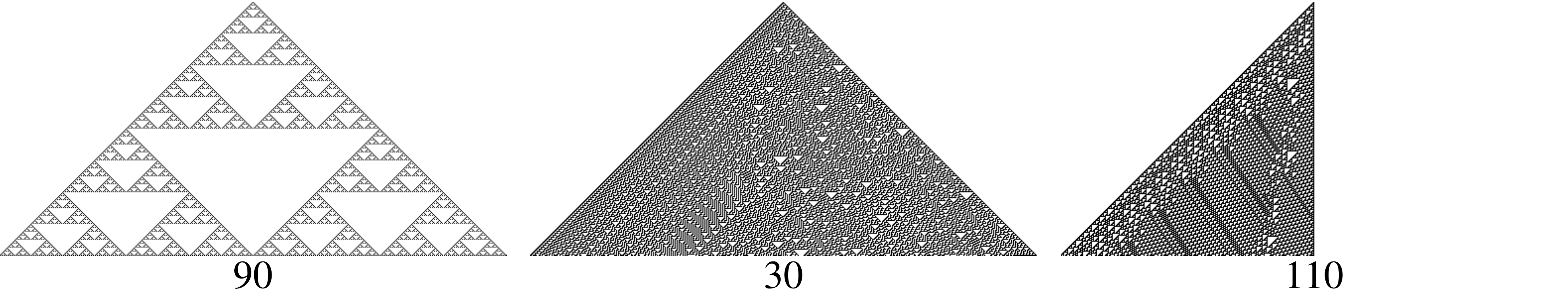}}
	\caption{
		Three $2$-color cellular automaton rules depending on $3$ cells, begun from a single black cell.
		Despite very different interior behavior, the boundaries all exhibit simple linear growth.
	}
	\label{rules_90,30,110}
\end{figure}

Our main purpose in this paper is to inventory the boundary growth of the $2^{16} = 65536$ one-dimensional rules that use $k = 2$ colors and that depend on $d = 4$ cells.
Several rules in this space have boundaries not found among rules with shorter range $d$. For example, some nested automata have piecewise linear boundaries characterized by morphic words, while more chaotic automata have boundaries that behave like random walks.

Boundaries of cellular automata have been studied before. Phillips~\cite{Phillips} studied the $k = 2$, $d = 3$ automata with periodic-background initial conditions, which is more general than the constant-background initial conditions that we consider here, and he found boundary growth rates that depend on the initial condition. In a live experiment in 2005, Wolfram~\cite{Wolfram live experiment} investigated the boundaries of $k = 2$, $d = 4$ rules begun from simple initial conditions. Our paper can be seen as the completion of this experiment.

Because of the large size of the rule space, we are particularly interested in making our inventory programmatically accessible so that it can be searched and computed with. The \emph{Mathematica} package \textsc{CellularAutomatonData}~\cite{CellularAutomatonData} provides an interface to all the data we accumulated both programmatically and by hand. The primary function in this package uses the same syntax as the data functions built into \emph{Mathematica}, and a cellular automaton is denoted $\{\{n, \, k, \, (d-1)/2\}, \, init\}$ to parallel \texttt{CellularAutomaton}.
For example,
\begin{verbatim}
	CellularAutomatonData[{{1273, 2, 3/2}, {{1}, 0}}, "GrowthRate"]
\end{verbatim}
retrieves the limiting growth rate of the $k = 2$, $d = 4$ rule number $1273$ begun from the initial condition $\singleblackcell$, which is $6/5$. The package \textsc{CellularAutomatonBoundaries}~\cite{CellularAutomatonBoundaries} contains code used to generate the data in \textsc{CellularAutomatonData}~\cite{CellularAutomatonData}. These packages are available from the websites of the authors~\cite{CellularAutomatonData,CellularAutomatonBoundaries}.

Section~\ref{background} of the paper establishes our notation and reviews the boundary growth rates for $2$-color cellular automata depending on at most $3$ cells. In Section~\ref{reducible} we describe a search for $2$-color cellular automata depending on $4$ cells that exhibit reducible boundary growth, and we discuss boundaries found by this search. In Section~\ref{irreducible} we address the automata that were not found to have reducible growth; we study their growth rates statistically using tools typically applied to random walks. We also attempt to assign a growth function $t^b$ to each automaton for some $0 \leq b \leq 1$. However, in Section~\ref{no exponent} we show that in general this is impossible, by constructing an automaton for which no such $b$ exists. In Section~\ref{conclusion} we discuss possible extensions and open questions.

Classifying automata by their boundaries identifies many automata with interesting behavior. Many boundaries closely reflect the behavior of the interior. For example, nested boundaries arise from nested automata, while chaotic boundaries arise from complex automata. Some automata with complicated interiors (such as rules 30 and 110 in Figure~\ref{rules_90,30,110}) nevertheless have simple boundaries. Thus the complexity of an automaton's boundary provides a lower bound on the complexity of its interior. Throughout the paper we describe many interesting automata found in this way by using the boundary as a filter.

\section{Background}\label{background}

\subsection{One-dimensional cellular automata}

The cellular automata that we study in this paper are one-dimensional.
A \emph{one-dimensional cellular automaton} consists of
\begin{itemize}
\item
an alphabet $\Sigma$ of size $k$,
\item
a positive integer $d$,
\item
a function $i$ from the set of integers to $\Sigma$, and
\item
a function $f$ from $\Sigma^d$ ($d$-tuples of elements in $\Sigma$) to $\Sigma$.
\end{itemize}
The function $i$ is called the \emph{initial condition}, and the function $f$ is called the \emph{rule}.
We think of the initial condition as an infinite row of discrete cells, each assigned one of $k$ colors.
To evolve the cellular automaton, we update all cells in parallel, where each cell updates according to $f$, a function of $d$ cells in its vicinity on the previous step.

There are $k^{k^d}$ rules on $k$ colors depending on $d$ cells. We adopt the usual convention of naming a cellular automaton's rule by the number whose base-$k$ digits consist of the outputs of the rule under the $k^d$ possible inputs of $d$ cells, ordered reverse-lexicographically. For example, the $2$-color rule depending on $3$ cells that maps the 8 possible inputs according to the table
\[
\hbox{
	\tiny
	\begin{tabular}{|c|c|c|c|c|c|c|c|}
		\hline
		\icontriple{\bc}{\bc}{\bc} & \icontriple{\bc}{\bc}{\wc} & \icontriple{\bc}{\wc}{\bc} & \icontriple{\bc}{\wc}{\wc} & \icontriple{\wc}{\bc}{\bc} & \icontriple{\wc}{\bc}{\wc} & \icontriple{\wc}{\wc}{\bc} & \icontriple{\wc}{\wc}{\wc} \\
		\iconsinglec{\wc} & \iconsinglec{\wc} & \iconsinglec{\wc} & \iconsinglec{\bc} & \iconsinglec{\bc} & \iconsinglec{\bc} & \iconsinglec{\bc} & \iconsinglec{\wc} \\
		\hline
	\end{tabular}
}
\]
is rule $00011110_2 = 30$ in this numbering. Here we have identified $0 = \wc$ and $1 = \bc$.

The evolution of a one-dimensional cellular automaton can be visualized in two dimensions by displaying each row below its predecessor. For example, Figure~\ref{rules_90,30,110} shows $2^8$ steps of three rules evaluated from the initial condition $\singleblackcell$. To create such pictures we must choose a \emph{horizontal offset}. For instance, the offset of rule $30$ in the table above is center-aligned: every cell depends on the cells in the same position, $l=1$ position to the left, and $r=1$ position to the right. For a different offset, the rows in the automaton will be the same; each row simply shifts with respect to the row preceding it. In other words, shifting $l$ and $r$ to $l - \Delta$ and $r + \Delta$, respectively, only shears the two-dimensional picture. Therefore, for convenience we generally choose a horizontal offset that minimizes the total width of the region of interest.

\subsection{Row lengths}

We require that all but finitely many cells in the initial condition have the same color. Then each row has finite length, which we define as follows. If all cells in a row are the same color, the length of that row is $0$. Otherwise, the length of a row is the number of cells in the region bordered by, and including, the first and last cells that differ from the constant background. For a given cellular automaton, let $\ell(t)$ be the length of the row on step $t$ for each $t \geq 0$.

For example, the length $\ell(t)$ for rule~$90$ begun from $\singleblackcell$ as in Figure~\ref{rules_90,30,110} is $\ell(t) = 2 t + 1$ for all $t \geq 0$. For rule~$30$ the length is also $\ell(t) = 2 t + 1$, whereas for rule~$110$ it is $\ell(t) = t + 1$. Note that $\ell(t)$ does not depend on the horizontal offset chosen to display the automaton.

At each step in a cellular automaton, information can propagate at most $l$ steps from the right boundary and at most $r$ steps from the left boundary, where $l$ and $r$ depend on the offset chosen but are subject to $l + 1 + r = d$. In other words, the maximum growth rate possible (called the ``speed of light'') is $d - 1$ cells per step, and if the maximum growth rate persists over time, then $\ell(t) = (d - 1) t + c$ for some $c$. If the maximum growth is achieved at every step, then $\ell(t) = (d - 1) t + \ell(0)$ for all $t \geq 0$.

Because each row in a cellular automaton depends only on the previous row, the difference sequence $\ell(t + 1) - \ell(t)$ is particularly relevant, since it gives the number of cells by which the automaton grows or shrinks at each step. It will be useful to think of the difference sequence as an infinite word on the set of integers.
\begin{definition}
The \emph{boundary word} of a cellular automaton is the sequence $\{\ell(t + 1) - \ell(t)\}_{t \geq 0}$.
\end{definition}
We will see that the boundary word frequently reflects properties of an automaton.

If the boundary word is eventually periodic, then $\ell(t)$ can be written as a piecewise expression in linear functions. Namely, there exist integers $m, t_\text{min}$ and rational numbers $a$ and $c_0, c_1, \dots, c_{m-1}$ such that for all $t \geq t_\text{min}$ we have
\begin{equation}\label{periodic boundary word}
	\ell(t) =
	\begin{cases}
		a t + c_0		& \text{if $t \equiv 0 \mod m$} \\
		a t + c_1		& \text{if $t \equiv 1 \mod m$} \\
		\quad\, \vdots		& \qquad \vdots \\
		a t + c_{m-1}		& \text{if $t \equiv m-1 \mod m$.}
	\end{cases}
\end{equation}
For example, the sequence $\ell(t)$ for rule~$45$ begun from $\singleblackcell$, shown in Figure~\ref{rules_45,107,106}, is $1, 3, 4, 6, 7, \dots$. The boundary word $212121\cdots$ is periodic with period length $2$, and the length of the row at step $t$ is
\[
	\ell(t) =
	\begin{cases}
		3 t / 2 + 1	& \text{if $t \equiv 0 \mod 2$} \\
		3 (t + 1)/ 2	& \text{if $t \equiv 1 \mod 2$}
	\end{cases}
\]
for $t \geq 0$. Rule~$107$ begun from a single black cell is also shown in Figure~\ref{rules_45,107,106}; its boundary word $1212020202\bar{4}202\bar{4}\cdots$, where $\bar{4} = -4$, is eventually periodic with period length $4$, and for $t \geq 7$
\[
	\ell(t) =
	\begin{cases}
		11	& \text{if $t \equiv 0 \mod 4$} \\
		11	& \text{if $t \equiv 1 \mod 4$} \\
		13	& \text{if $t \equiv 2 \mod 4$} \\
		9	& \text{if $t \equiv 3 \mod 4$.}
	\end{cases}
\]

\begin{figure}
	\scalebox{.3}{\includegraphics{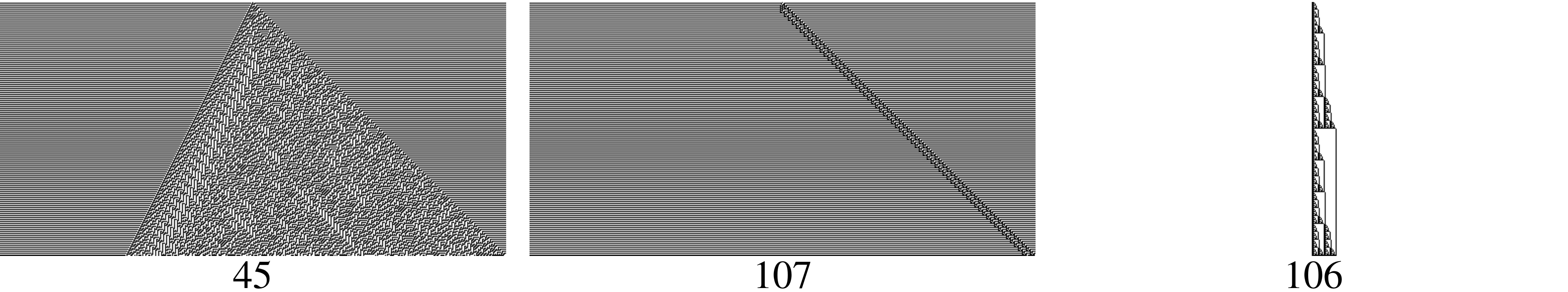}}
	\caption{
		Rules~$45$ and $107$ have row lengths that can be expressed by Equation~\ref{periodic boundary word}.
		Rule~$106$ exhibits square-root growth when begun from two adjacent black cells.
	}
	\label{rules_45,107,106}
\end{figure}

All $2$-color cellular automata depending on $d = 2$ cells have eventually periodic boundary words, either with growth rate $a = 0$ or $a = 1$.

The boundaries of $2$-color cellular automata depending on $d = 3$ cells are largely similar.
These automata generate a variety of internal structures: rule $90$, for example, produces nested structure, while rules $30$ and $110$ yield complex behavior. One new feature seen for $d = 3$ is square-root growth, exhibited for example by rule~$106$ begun from the initial condition $\twoblackcells$, as shown in Figure~\ref{rules_45,107,106}. We discuss square-root growth further in Section~\ref{square-root growth}. However, with the exceptions of rules~$106$, $120$, $169$, and $225$, each $2$-color cellular automaton depending on $d = 3$ cells has an eventually periodic boundary word. Moreover, for every automaton in this space (with a constant-background initial condition), the limiting growth rate
\begin{align}
\label{limitinggrowthrate}
	\lim_{t \to \infty} \frac{\ell(t)}{t}
\end{align}
exists and is an element of $\{0, 1, 3/2, 2\}$.
In particular, for rule~$106$ this limit is $0$.
(Note that if we allow a general periodic background for the initial condition, then the boundary word is not necessarily eventually periodic; for example, the left boundary of rule~$30$ begun from the initial condition $\cdots\wc\s\bc\s\wc\s\bc\s\wc\s\bc\s\bc\s\bc\s\wc\s\bc\s\wc\s\bc\s\wc\cdots = \cdots0101011101010\cdots$ appears to be chaotic.)

In general, the limiting growth rate $\lim_{t \to \infty} \ell(t)/t$ of a cellular automaton may not exist, as we see in Section~\ref{reducible}. Moreover, the limiting growth exponent $\lim_{t \to \infty} \log_t \ell(t)$ may not exist, as we show in Section~\ref{no exponent}. However, in most cases these values do appear to exist, so in Section~\ref{irreducible} we use them as statistical information about boundaries.

We mention the observation of Phillips~\cite{Phillips} that if the sequence of rows in an automaton is not eventually periodic, then $\ell(t)$ grows at least logarithmically.
This is because for $\ell \geq 2$ there are $k^{\ell - 1} (k - 1)^2$ possible rows of length $\ell$, so a cellular automaton that never returns to the same state has at most exponentially many rows of length $\ell$. Logarithmic growth is not seen for $k = 2$ and $d \leq 3$, and we did not find logarithmic growth among $d = 4$ rules either. However, it is possible to construct an automaton that implements counting in binary, by using additional colors (and additional steps) to propagate carries, and this automaton grows logarithmically~\cite{Phillips}.

\section{Automata with reducible growth}\label{reducible}

In this section we describe a combined automated--manual search for reducible boundaries among all $2$-color rules depending on $d = 4$ cells, begun from single-cell initial conditions. Eventually periodic boundary words can be detected completely automatically, and we examine by hand the automata that are not found to have an eventually periodic boundary word.

As in every space of cellular automaton rules, some rules in this space are equivalent to others by simple transformations. For example, reversing each tuple in the definition of the rule and reversing the initial condition results in an image that is simply the left--right reflection of the original. Similarly, permuting the colors in a rule and in the initial condition produces an image that is obtained from the original by the same permutation. Therefore it suffices to consider only one rule among each equivalence class of rules obtained by reflecting and permuting. For this we choose the rule with minimal rule number. For $k=2$ and $d=4$ this reduces the number of rules from $65536$ to $16704$.

As a simplifying assumption, we consider only the two initial conditions $\singleblackcell$ and $\singlewhitecell$, each consisting of an infinite constant background with a single perturbed cell. This results in $33408$ automata (two initial conditions for each rule). In many cases these initial conditions suffice to characterize the growth of the rule. However, for rules that grow dramatically differently depending on the initial condition, the data we collect may not be representative of typical growth.

We further restrict the initial condition by requiring the background color to reoccur on some later step (but not necessarily the next step). That is, we only consider the initial condition $\singleblackcell$ if a white background reoccurs on some later step. Similarly, we only consider $\singlewhitecell$ if a black background reoccurs on a later step. We ignore these initial conditions otherwise because we are interested in long-term behavior, and a background that does not reoccur is a type of transience. Doing so reduces the number of automata to $25088$.

We run each of these automata for $t_\text{max}$ steps and consider the difference sequence $\ell(t+1) - \ell(t)$ for $t_\text{min} \leq t \leq t_\text{max} - 1$, with some $t_\text{min} > 0$ allowing for transience. Let $m$ be the smallest positive integer such that $\ell(t+1+m) - \ell(t+m) = \ell(t+1) - \ell(t)$ for all $t_\text{min} \leq t \leq t_\text{max} - 1 - m$. If $m < (t_\text{max} - t_\text{min})/4$, then we deem the boundary word to be eventually periodic (and $\ell(n)$ to satisfy Equation~\ref{periodic boundary word}), and we record the period length $m$ and the growth rate
\begin{equation}\label{record growth rate}
	a = \frac{\text{sum of the terms in the period}}{m} = \frac{\ell(t+m) - \ell(t)}{m}.
\end{equation}
Otherwise we consider the period length unreliable and this test inconclusive.

In choosing a time range $t_\text{min} \leq t \leq t_\text{max}$ on which to test periodicity, we face opposing goals: To overcome possible transience, we want $t_\text{min}$ to be large, but for speed we want $t_\text{max}$ to be small. Our solution is to use the four short time ranges $20 \leq t \leq 100$, $50 \leq t \leq 300$, $200 \leq t \leq 600$, and $400 \leq t \leq 1000$ as successive filters, followed by a more extensive range. If a reliable period length is found in any of these ranges, then we skip the remaining ranges and compute the period length in a final range $500 \leq t \leq 4000$ to confirm that the period length persists.
This final time range identifies only $32$ corrections to period lengths found by one of the first four ranges, and all but one of those (correcting the slope from $7/4$ to $18/11$ for rule~$23726$ begun from $\singleblackcell$) are cases where the boundary word does not appear to be eventually periodic. Running all $25088$ automata through the four filter ranges took approximately twenty minutes on a 2.5GHz machine. Running the final time range took approximately two and a half days.

While we believe that confirming each period length in the range $500 \leq t \leq 4000$ has allowed our data to be highly reliable, this algorithm clearly does not guarantee that if a period length was found then the boundary word is in fact eventually periodic (false positives), nor does it guarantee that if a period length was not found then the boundary word is not eventually periodic (false negatives). There are several automata whose boundaries do not stabilize until well after $400$ steps or whose eventual behavior is unclear. For example, rule~$11109$ begun from $\singleblackcell$ grows to $\ell(1722) = 918$, and thereafter has average growth rate $0$. Rule~$4713$ begun from $\singleblackcell$ jettisons a particle at step~$915$, leaving behind an otherwise chaotic left boundary. Rule~$10633$ (begun from either initial condition) appears to have an eventually periodic boundary word due to its internal froth generally moving away from the boundary, but it is not clear that this will continue indefinitely. Worse, there are automata whose growth is periodic for short time intervals but that are most likely not periodic in general. For example, rule~$457$ begun from either initial condition has a boundary word that is periodic in the range $100 \leq t \leq 200$; but for larger ranges we see that the periodicity does not continue.

These examples indicate that in general one cannot determine the long-term behavior of the boundary of a cellular automaton by examining finitely many steps. Of course, this is not surprising, because the boundary can depend sensitively on the interior of the automaton, and it is known that some cellular automaton rules are computationally universal. Indeed, we determined the four time ranges only after some experimentation with a selection of rules.

Executing this automatic search yields $837$ automata (with $620$ distinct rules) that were not found to have an eventually periodic boundary word. Among these $837$, there are only $757$ distinct pictures (at least for $500$ rows), because several pairs of inequivalent rules appear to nonetheless generate the same evolution due to certain configurations not appearing. We examined each of these classes manually and found that $36$ automata do in fact appear to have eventually periodic boundary words, while another $81$ exhibit self-similarity. Therefore a classification of the $25088$ automata is as follows.
\begin{enumerate}
\item
$24287$ automata have eventually periodic boundary words.
\item
$81$ automata have boundary words that are not eventually periodic but are reducible.
\item
$720$ automata have boundaries that are most likely not reducible.
\end{enumerate}
Analyzing automata in the third class is the subject of Section~\ref{irreducible}. Automata in the first two classes have boundary words with simple descriptions, and they are the subject of this section.

A note regarding the level of rigor is in order. We do not formally prove the claims in this section, neither the explicit growth rates nor other properties we describe. Therefore they can either be taken as conjectures or as semi-rigorous results that are experimentally verified for the first $4000$ (and in some cases many more) steps of the cellular automata involved. Proving each claim is beyond the scope of the current paper, although we touch on this in Section~\ref{conclusion}.

\subsection{Eventually periodic boundary words}\label{Eventually periodic boundary words}

In the first class of $24287$ automata---those with eventually periodic boundary words---the most common average growth rate is $0$, and there are $11768$ automata with growth rate $0$. The following table gives the thirty most common growth rates $a$ (as in Equation~\ref{periodic boundary word}) and the number $N_a$ of automata with each rate.
\begin{center}
	\begin{tabular}{cc|cc|cc|cc|cc}
		$a$ & $N_a$ & $a$ & $N_a$ & $a$ & $N_a$ & $a$ & $N_a$ & $a$ & $N_a$ \\ \hline
		0 & 11768 & 5/4 & 102 & 15/13 & 18 & 9/7 & 11 & 8/7 & 8 \\
		3 & 4800 & 5/3 & 73 & 9/4 & 17 & 10/7 & 10 & 13/8 & 7 \\
		2 & 4001 & 6/5 & 53 & 9/5 & 17 & 7/6 & 10 & 11/10 & 7 \\
		1 & 1082 & 7/4 & 45 & 7/5 & 17 & 15/14 & 10 & 14/11 & 6 \\
		5/2 & 985 & 3/4 & 40 & 5/6 & 15 & 1/2 & 10 & 7/8 & 6 \\
		3/2 & 951 & 4/3 & 28 & 11/8 & 11 & 9/8 & 9 & 2/3 & 6
	\end{tabular}
\end{center}

The smallest nonzero growth rate is $2/5$, and $5$ automata have growth rate $2/5$.

If $r/s$ is a non-negative rational number written in lowest terms (with $\gcd(r,s) = 1$), let us define the \emph{height} of $r/s$ to be $\max(r, s)$. The height of a number is one measure of its complexity.
The automaton whose limiting growth rate has largest height is rule~$10168$ begun from a single black cell, with growth rate $a = 1578/1013$. This automaton also has the largest period length: $2026$ steps. The next largest-height growth rates that occur are $773/411$, $515/318$, $398/247$, $329/199$, and $297/127$; all these automata have fairly simple interiors.

The growth rate with largest height that is generated by two distinct rules (not two distinct \emph{automata} that share the same rule but two distinct \emph{rules}) is $91/55$. The rules are $17380$ and $46236$, respectively begun from $\singleblackcell$ and $\singlewhitecell$.

\subsection{Morphic words}

The remainder of Section~\ref{reducible} concerns automata in the second class mentioned above: automata with boundary words that are not eventually periodic but still amenable to short description. These automata all have boundaries that exhibit nontrivial self-similarity, so we may refer to these as fractal boundaries. It turns out that the boundary words for all these automata are \emph{morphic words}---words generated by iterating a morphism (also known as a substitution system).

Let $\Sigma$ and $\Delta$ be finite alphabets, and let $\Sigma^*$ denote the set of all finite words with letters in $\Sigma$. The empty word is denoted by $\epsilon$. For a function $\varphi : \Sigma \to \Delta^*$ and a (finite or infinite) sequence $w_0, w_1, \dots$ of letters in $\Sigma$, define $\varphi(w_0 w_1 \cdots) = \varphi(w_0) \varphi(w_1) \cdots$. We refer to $\varphi$ as a \emph{morphism}, since $\varphi(x y) = \varphi(x) \varphi(y)$ for all words $x, y$. If $\Delta = \Sigma$ and there is some letter $A \in \Sigma$ and some word $x \in \Sigma^*$ such that $\varphi(A) = A x$, then by iteratively applying $\varphi$ to $A$ we obtain prefixes of the word
\[
	\varphi^\omega(A) \colonequal A \, x \, \varphi(x) \, \varphi^2(x) \, \cdots,
\]
which is a fixed point of $\varphi$. Moreover, this word is the unique fixed point of $\varphi$ beginning with $A$. An infinite word (or, equivalently, an infinite sequence) $\textbf{w}$ is \emph{morphic} if there is a letter $A \in \Sigma$ and morphisms $\varphi : \Sigma \to \Sigma^*$ and $\psi : \Sigma \to \Delta^*$ such that
\[
	\textbf{w} = \psi(\varphi^\omega(A)).
\]
We see in the following subsections that, for each cellular automaton with reducible boundary structure, the boundary word is morphic (and is a word on some finite set $\Delta$ of integers).

In the next three subsections we address fractal automata whose limiting growth rates exist. We will see that these rates do not approach the complexity of some of the growth rates observed for eventually periodic boundary words in Section~\ref{Eventually periodic boundary words}. In the final subsection we discuss automata whose limiting growth rates do not exist. (Many of the rules discussed have nearly identical behavior when begun from the two initial conditions; in these cases we only discuss one initial condition without mentioning this further.)

We refer the reader to the book of Allouche and Shallit~\cite[Chapters~6--8]{Automatic Sequences} for a comprehensive treatment of morphic words. For our immediate purposes, it suffices to mention that prepending a word to a morphic word produces another morphic word. In particular, every eventually periodic word is morphic.

\subsection{Square-root growth}\label{square-root growth}

Before discussing square-root growth among $2$-color rules depending on $4$ cells, we first discuss $d=3$ rule~$106$, which also exhibits square-root growth. Figure~\ref{rules_45,107,106} shows the evolution of rule~$106$ begun from two adjacent black cells. The boundary word of this automaton is the infinite word
\[
	\textbf{w}_{106} = 11010011000000010000000011010011\cdots
\]
on the alphabet $\{0, 1\}$. Let us rewrite the boundary word as
\[
	\textbf{w}_{106} = 1^2 0^1 1^1 0^2 1^2 0^7 1^1 0^8 1^2 0^1 1^1 0^2 1^2 0^{31} 1^1 0^{32} \cdots,
\]
since the run lengths of each block suggest a pair of morphisms that generate $\textbf{w}_{106}$. In particular, observe that replacing each $0$ in $\textbf{w}_{106}$ by $0^4$ causes $0^8$ to become $0^{32}$. So that $0^7 1^1 0^8 \to 0^{31} 1^1 0^{32}$, we need $1 \to 0001$; however, not every $1$ can be replaced using this rule, since this would result in no instances of $1^2$ in the fixed point. Therefore we introduce some additional letters. Consider the morphism
\[
	\varphi =
	\{
		A \to ABCD, \;
		B \to CCAB, \;
		C \to CCCC, \;
		D \to CCCD
	\}.
\]
The fixed point $\varphi^\omega(A)$ of this morphism is
\[
	A^1 B^1 C^1 D^1 C^2 A^1 B^1 C^7 D^1 C^8 A^1 B^1 C^1 D^1 C^2 A^1 B^1 C^{31} D^1 C^{32} \cdots.
\]
Applying the morphism $\psi = \{A \to 1, B \to 1, C \to 0, D \to 1\}$ to this fixed point gives $\textbf{w}_{106} = \psi(\varphi^\omega(A))$.

From this morphism one can derive that rule~$106$ grows like $\sqrt{t}$. Here we show a weaker claim---that $1/2$ is a limit point of the sequence $\log_t \ell(t)$. Letting $E = ABCDCCAB$ and $F_n = C^{2\cdot 4^n-1}DC^{2\cdot4^n}$, one can check that
\[
\phi^\alpha(A) = \bigg (\prod_{k=1}^{2^{\alpha-2}-1} E F_{\nu_2(k)+1} \bigg ) E C^{2^{2\alpha-1}-1}D \qquad \text{for $\alpha \geq 3$},
\]
where $\nu_2(k)$ is the exponent of the highest power of $2$ dividing $k$. Using $\nu_2(k)$ to count occurrences of $E$ and $F_k$ in $\phi^\alpha(A)$ preceding $C^{2^{2\alpha-1}-1}D$ gives
\[
\ell(2^{2\alpha-1}) = \ell(0) + \sum_{t=0}^{2^{2\alpha-1}} \textbf{w}_{106}(t) = 3\cdot2^{\alpha-1} +1 \qquad \text{for $\alpha \geq 1$}.
\]
This agrees with the observation of Gravner and Griffeath~\cite{RPS} that the configuration at step $2^{2\alpha-1}$ is
\[
	\cdots\wc\s\wc\s\wc\s\bc\s\bc\s\underbrace{\s\wc\s\wc\cdots\wc\s\wc\s}_{3 \cdot 2^{\alpha-1} - 2}\s\bc\s\wc\s\wc\s\wc\cdots.
\]
Including the trailing $C^{2^{2\alpha-1}-1}D$ in $\phi^\alpha(A)$ leads to $\ell(2^{2\alpha}) = 3\cdot2^{\alpha-1}+2$ for $\alpha \geq 1$.

Among $2$-color rules depending on $4$ cells, two inequivalent rules exhibit square-root growth from single-cell initial conditions: rules $34394$ and $39780$. Although they are not equivalent as rules, the automata obtained by running these rules from a single black cell are equivalent under left--right reflection, since the tuple on which the rules differ does not occur in the evolution begun from a single black cell. In particular, rule~$39780$ is known to exhibit conditional reversibility, due to the local rule being a bijective function in the leftmost position~\cite{Rowland 2006}, whereas rule~$34394$ does not have this property. Figure~\ref{rules_39780,3701,7195,8067,27898} shows rule~$39780$.

For both of these automata, the boundary word is
\[
	\textbf{w}_{39780} = 2210221\bar{1}1\bar{1}1\bar{1}102210221\bar{1}1\bar{1}1\bar{1}1\bar{1}\cdots,
\]
a word on the alphabet $\{-1, 0, 1, 2\}$, where we have written $\bar{1}$ for $-1$. Because of the repeating $1 \bar{1}$ oscillations, the run lengths of the original sequence do not reveal much. However, partitioning into blocks of length $2$ as
\begin{multline*}
	\textbf{w}_{39780} = (2 2)^1 (1 0)^1 (2 2)^1 (1 \bar{1})^3 (1 0)^1 (2 2)^1 (1 0)^1 (2 2)^1 (1 \bar{1})^{15} \\
		(1 0)^1 (2 2)^1 (1 0)^1 (2 2)^1 (1 \bar{1})^3 (1 0)^1 (2 2)^1 (1 0)^1 (2 2)^1 (1 \bar{1})^{63} \cdots
\end{multline*}
shows some structure.
If $\varphi$ is the morphism
\begin{multline*}
	\{
		A \to ABC, \;
		B \to DAB, \\
		C \to CECE, \;
		D \to CECD, \;
		E \to CECE
	\}
\end{multline*}
and $\psi = \{A \to 2, B \to 2, C \to 1, D \to 0, E \to \bar{1}\}$, then $\textbf{w}_{39780} = \psi(\varphi^\omega(A))$. To show again that $1/2$ is a limit point of $\log_t \ell(t)$, let $F = DABCDABC$ and $G(n) = (EC)^n$. Then for $\alpha \geq 3$
\[
D \phi^\alpha(A) =  \bigg (\prod_{k=1}^{2^{\alpha-2}-1} F G(2^{2\nu_2(2k)}-1) \bigg ) F G(2^{2\alpha -3}-1) E,
\]
where again $\nu_2(k)$ is the exponent of the highest power of $2$ dividing $k$. Counting occurrences of $F$ and $G(n)$ in $D\phi^\alpha(A)$ and computing their respective lengths and contributions to the boundary, we obtain
\[
\ell \big ( 4^{\alpha-1/2} +2^{\alpha-1} \big ) = 5 \cdot 2^{\alpha-1} \qquad \text{for $\alpha \geq 2$}.
\]

\begin{figure}
	\begin{center}
		\scalebox{.25}{\includegraphics{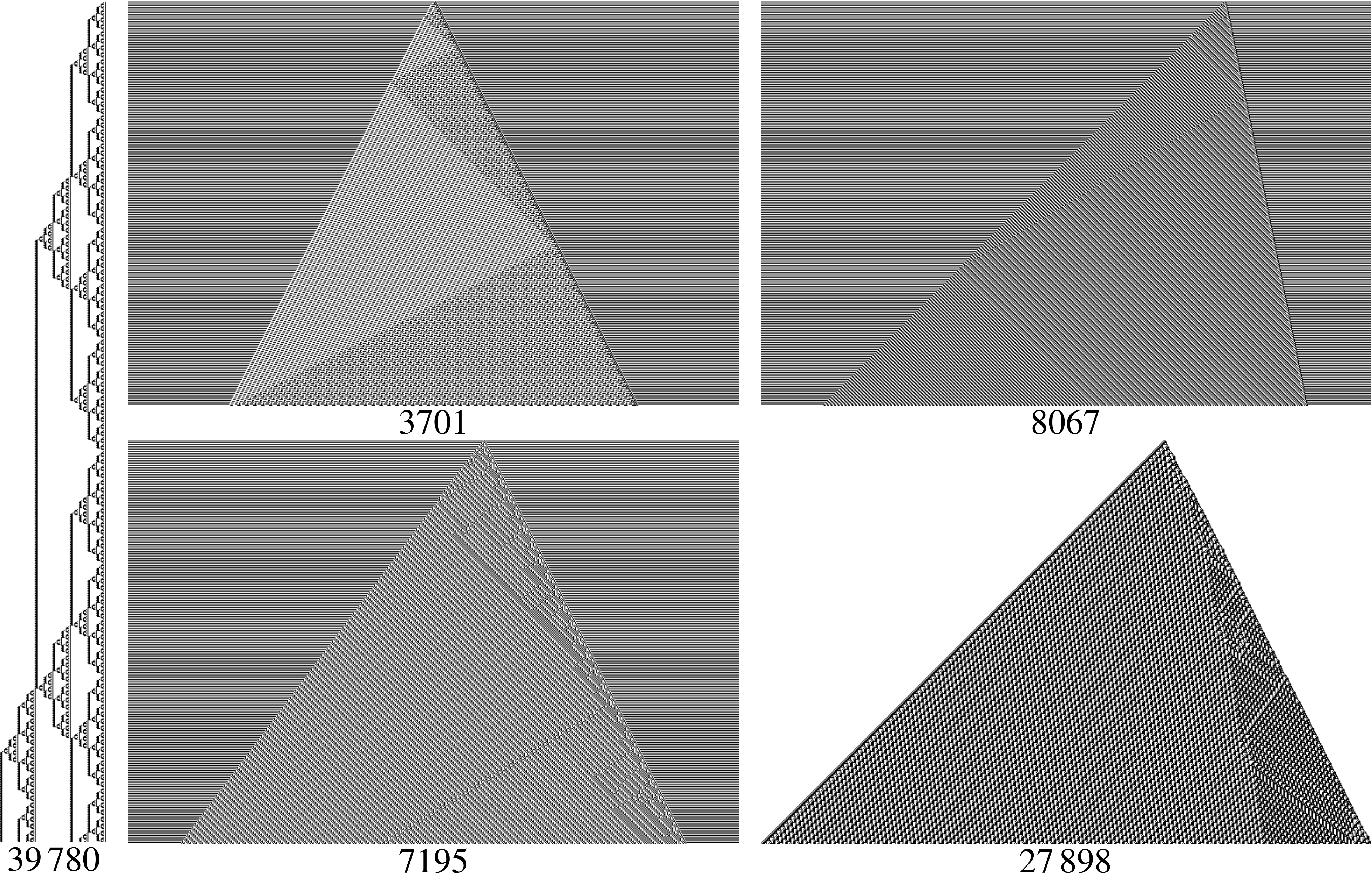}}
	\end{center}
	\caption{
		Rule~$39780$ grows like $\sqrt{t}$.
		The other four automata pictured here contain oscillating particles.
	}
	\label{rules_39780,3701,7195,8067,27898}
\end{figure}

Note that the morphism $\varphi$ for rule~$106$ is $4$-uniform. Consequently, the sequence $\textbf{w}_{106}$ is $2$-automatic (meaning that there is a finite automaton that outputs the $t$th term when input the binary digits of $t$); it follows that $\ell(t)$ is $2$-regular in the sense of Allouche and Shallit~\cite{Allouche--Shallit 1992, Allouche--Shallit 2003} and therefore can be computed quickly. On the other hand, the morphism $\varphi$ for rule~$39780$ is not uniform, and indeed it appears that the sequence $\ell(t)$ for this automaton is not $k$-regular for any small value of $k$.

\subsection{Oscillating particles}

Four rules have boundary words that are nearly periodic but that are perturbed occasionally by particles that oscillate in the interior of the automaton. They are shown in Figure~\ref{rules_39780,3701,7195,8067,27898}.

First consider rule~$3701$ begun from a single black cell. The right boundary is not perturbed when the particle reflects off of it, but the left boundary is perturbed at steps $(3 \cdot 5^\alpha + 5)/4 - \alpha$ for $\alpha \geq 0$. However, since the step numbers of these perturbations decay exponentially, they do not impact the limiting growth rate, so the limiting growth rate is $1$. The boundary word is generated from $A$ by the morphism $\varphi = \{A \to AB, B \to BC^6, C \to C^5\}$ followed by $\psi = \{A \to \epsilon, B \to 30, C \to 3\bar{1}\}$, where $\bar{1} = -1$.

Rule~$8067$ begun from a single black cell is similar, with a single particle perturbing the left boundary at steps $(8 \cdot 7^{\alpha + 1} + 6 \alpha - 2)/9$. However, the particle also perturbs the right boundary when it reflects at steps $(20 \cdot 7^\alpha + 6 \alpha + 79)/9$.

Rule~$7195$ begun from a single black cell contains additional internal structures, but the net effect is that a single particle oscillates between the left and right boundary, with the rest of the structure remaining close to the right boundary. The particle in fact does not perturb the right boundary when it reflects, although it does perturb the left boundary.

Rule~$27898$ begun from a single black cell differs in two ways from the others. The oscillating particle does not traverse the entire interior width of the automaton but reflects off an internal boundary. Additionally, the ``particle'' at times looks more like a group of particles, and not every interaction with the boundary is identical. However, after four reflections the particle returns to its original state, so the oscillatory behavior is in fact simple.

The respective limiting growth rates for rules~$8067$, $7195$, and $27898$ are $6/5$, $5/4$, and $3/2$.
Although we do not determine the morphisms here, the regularity of the oscillations in these automata imply that the boundary words are morphic.

\subsection{Two automata with limiting growth rates}

Figure~\ref{rules_1273,36226} shows rule~$1273$ begun from a single black cell and rule~$36226$ begun from a single white cell. The boundary words for these automata are not eventually periodic, but they are morphic.  Moreover, the limiting growth rate $\lim_{t \to \infty} \ell(t)/t$ (Equation~\ref{limitinggrowthrate}) exists for each.

We begin with rule~$36226$ because its boundary is simpler. On a global scale this automaton exhibits nested structure similar to that produced by $d = 3$ rule~$90$ begun from a single black cell (see Figure~\ref{rules_90,30,110}). However, the right boundary is fractal.
The boundary word 
\[
	\textbf{w}_{36226} = 12211221221112212211221221111221 \cdots.
\]
can be obtained by dropping the first two letters in the fixed point $2212211\cdots$ of the morphism $\varphi = \{1 \to 1, 2 \to 221\}$. Recalling that $\nu_2(n)$ denotes the exponent of the highest power of $2$ dividing $n$, we can also write
\[
	\textbf{w}_{36226} = \prod_{n \geq 2} 1^{\nu_2(n)} 2.
\]
The limiting growth rate of the automaton is determined by the frequencies of $1$ and $2$ in $\textbf{w}_{36226}$.
The \emph{frequency} of a letter $x$ in an infinite word $w_0 w_1 \cdots$ is
\[
	\lim_{t \to \infty} \frac{|\{ 0 \leq i \leq t - 1 : w_i = x \}|}{t}.
\]
To compute the letter frequencies, we examine the incidence matrix of $\varphi$, which records for each pair of letters $x, y$ the number of occurrences of $x$ in $\varphi(y)$.
The incidence matrix for $\varphi$ is
\[
	\begin{bmatrix}
	 1 & 1 \\
	 0 & 2
	\end{bmatrix}.
\]
If the frequency of each letter in a morphic word $\varphi^\omega(A)$ exists, then the vector whose components are the letter frequencies is an eigenvector of the incidence matrix corresponding to the largest positive eigenvalue~\cite[Theorem~8.4.6]{Automatic Sequences}.
In the case of $\textbf{w}_{36226}$, the letter frequencies exist, and that vector is $(1/2, 1/2)$.
Therefore the letters $1$ and $2$ occur with equal frequency, and on average the automaton grows $3/2$ cells per step.

\begin{figure}
	\begin{center}
		\scalebox{1}{\includegraphics{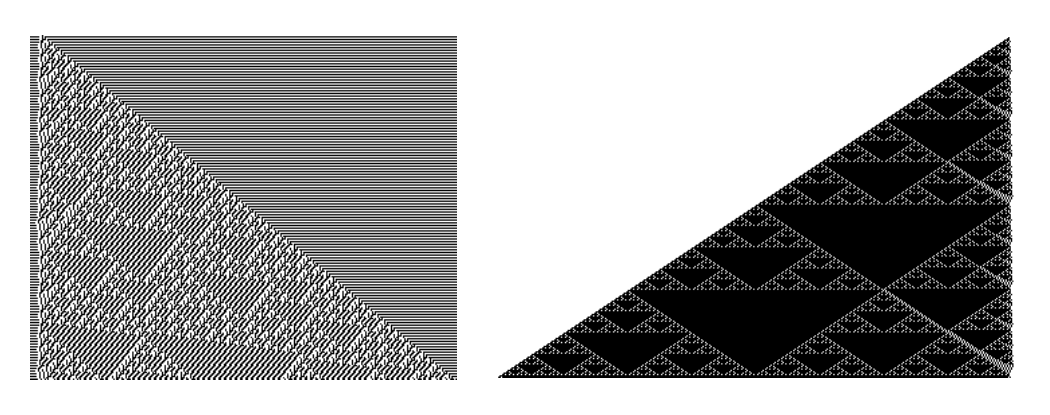}}
	\end{center}
	\caption{Rows $0$ through $2^8 - 1$ of rule~$1273$ and rule~$36226$, where the limiting growth rates have been used to shear the images such that the nonperiodic boundaries are vertical. The colors of rule~$36226$ have been reversed to place it against a white background.}
	\label{rules_1273,36226}
\end{figure}

Now consider rule~$1273$ begun from a single black cell.
The interior is also nested, although the nestedness is not as obvious visually.
For this automaton, the boundary word
\[
	\textbf{w}_{1273} = 3 \bar{1} 3 0 3 0 3 \bar{1} 3 \bar{1} 3 \bar{1} 3 0 3 0 3 \bar{1} 3 \bar{1} 3 \bar{1} 3 0 3 0 3 \bar{1} 3 0 3 0 \cdots
\]
(where again $\bar{1} = -1$) is given by $(3\bar{1})^1 (30)^2 \psi(\varphi^\omega(A))$, where
\[
	\varphi =
	\{
	A \to AC, \;
	B \to AD, \;
	C \to BA, \;
	D \to BB
	\},
\]
and $\psi$ maps
\begin{align*}
	A &\to (3\bar{1})^3 (30)^2 (3\bar{1})^3 (30)^2 (3\bar{1})^1 (30)^2 \\
	B &\to (3\bar{1})^3 (30)^2 (3\bar{1})^5 (30)^2 \\
	C &\to (3\bar{1})^5 (30)^2 (3\bar{1})^3 (30)^2 (3\bar{1})^1 (30)^2 \\
	D &\to (3\bar{1})^5 (30)^2 (3\bar{1})^5 (30)^2.
\end{align*}
The incidence matrix for $\varphi$ is
\[
	\begin{bmatrix}
	 1 & 1 & 1 & 0 \\
	 0 & 0 & 1 & 2 \\
	 1 & 0 & 0 & 0 \\
	 0 & 1 & 0 & 0
	\end{bmatrix},
\]
so the vector with components equal to the frequencies of the four letters $A, B, C, D$ is $(4/9, 2/9, 2/9, 1/9)$. The letters $A$, $B$, $C$, and $D$ correspond to respective net changes of $32$, $28$, $36$, and $32$ cells over $26$, $24$, $30$, and $28$ steps, and so one computes that the limiting growth rate is $6/5$ cells per step.

\subsection{Automata with no limiting growth rate}

Finally, a number of automata have linear growth in the sense that the limiting growth exponent $\lim_{t \to \infty} \log_t \ell(t)$ is $1$ although the limiting growth rate $\lim_{t \to \infty} \ell(t)/t$ does not exist.

As a typical example, consider rule~$2230$ begun from a single black cell.
The boundary word is
\[
	\textbf{w}_{2230} = 2^1 0^1 2^3 0^2 2^6 0^4 2^{12} 0^8 2^{24} 0^{16} \cdots.
\]
Replacing $0 \to 00$ and $2 \to 22$ produces $\textbf{w}_{2230}$ again, with the exception of the first three letters $202$. In other words, the structure of $\textbf{w}_{2230}$ is that of the fixed point beginning with $A$ of the morphism $\varphi = \{A \to ABCB, B \to BB, C \to CC\}$:
\[
	\varphi^\omega(A) = A^1 B^1 C^1 B^3 C^2 B^6 C^4 B^{12} C^8 B^{24} C^{16} \cdots.
\]
Applying $\psi = \{A \to \epsilon, B \to 2, C \to 0\}$ produces $\textbf{w}_{2230}$.

The frequencies of the letters $B$ and $C$ in $\varphi^\omega(A)$ turn out to not exist:
The frequency of $B$ in the first $4 \cdot 2^\alpha - 2$ letters is $(3 \cdot 2^\alpha - 2)/(4 \cdot 2^\alpha - 2)$, and the frequency of $B$ in the first $5 \cdot 2^\alpha - 2$ letters is $(3 \cdot 2^\alpha - 2)/(5 \cdot 2^\alpha - 2)$.
Since $3/4$ and $3/5$ are both limit points of $|\{ 0 \leq i \leq t - 1 : w_i = B \}|/t$, the frequency of $B$ does not exist.
Similarly, the frequency of $C$ does not exist.

Consequently, the frequencies of $0$ and $2$ do not exist in $\textbf{w}_{2230}$, and the growth rate $\lim_{t \to \infty} \ell(t)/t$ does not exist. However, the growth can still be quantified by $a_\text{inf} \colonequal \liminf \ell(t)/t = 6/5$ and $a_\text{sup} \colonequal \limsup \ell(t)/t = 3/2$.

\begin{figure}
	\begin{center}
		\includegraphics{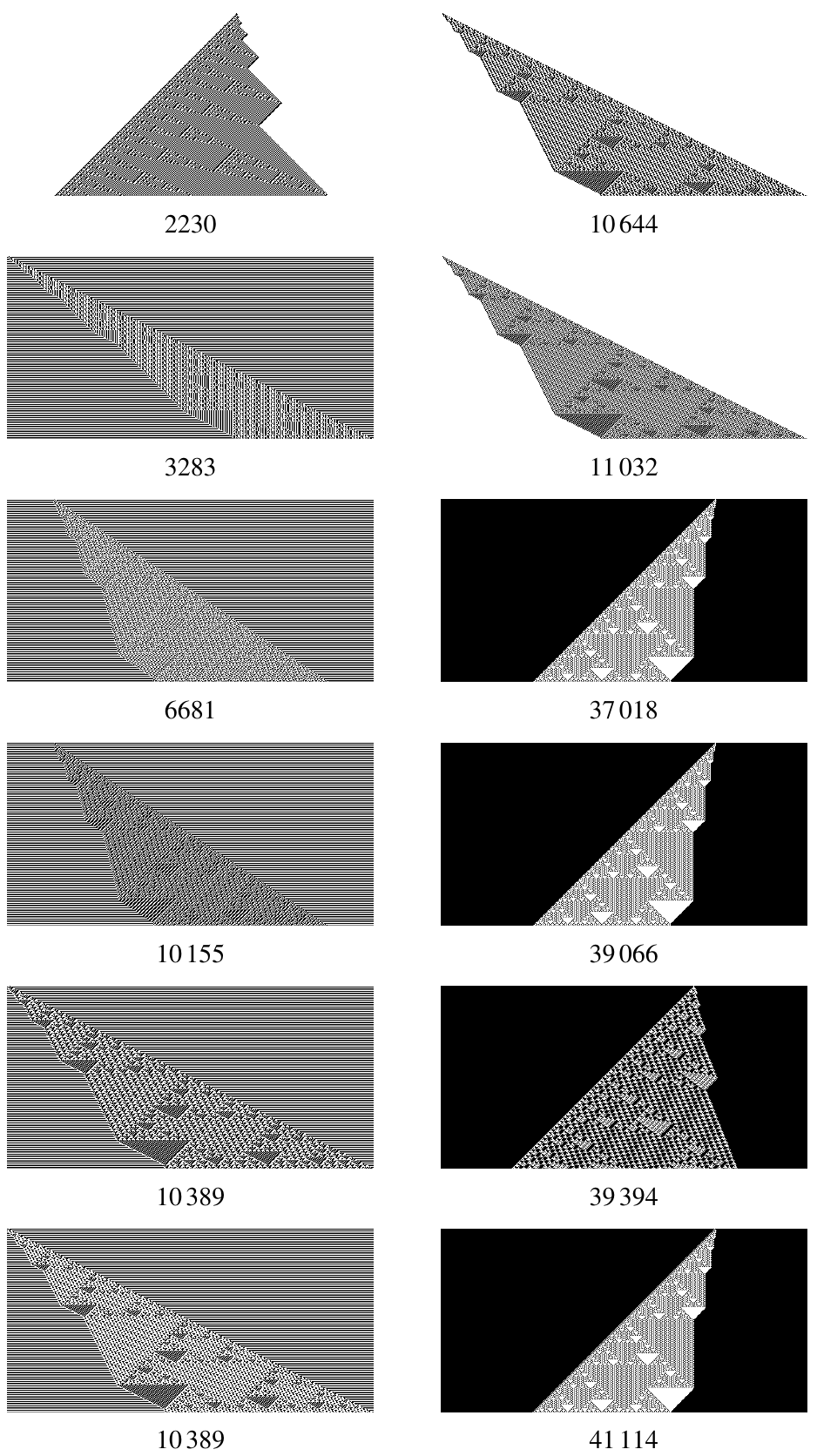}
	\end{center}
	\caption{
		Some nested automata with boundary words generated by the morphism $\{A \to ABCB, B \to BB, C \to CC\}$.
		They are variants on a common underlying structure, for which the limiting growth rate does not exist.
	}
	\label{nested_rules}
\end{figure}

Several other automata have boundaries that are also generated by the morphism $\varphi = \{A \to ABCB, B \to BB, C \to CC\}$, followed by some morphism $\psi$. The values $a_\text{inf}$ and $a_\text{sup}$ can be computed for these automata as well. Representatives are shown in Figure~\ref{nested_rules}, and bounds on their growth are given in the following table.

\[
\begin{array}{rcccccc}
 \text{rule} & \text{initial condition} & \psi(A) & \psi(B) & \psi(C) & a_\text{inf} & a_\text{sup} \\ \hline
 2230 & \singleblackcell & \epsilon & 2 & 0 & 6/5 & 3/2 \\
 3283 & \singleblackcell & 21 & 2\bar{1}21 & 2\bar{2} & 3/4 & 6/7 \\
 6681 & \singleblackcell & 31 & 3\bar{2}31 & 3\bar{3} & 15/16 & 15/14 \\
 10155 & \singlewhitecell & 2121 & 1121 & 1\bar{1} & 15/16 & 15/14 \\
 10389 & \singleblackcell & 32 & 3\bar{2}32 & 3\bar{3} & 9/8 & 9/7 \\
 10389 & \singlewhitecell & 230 & 3030 & 3\bar{3} & 9/8 & 9/7 \\
 10644 & \singleblackcell & 2 & 12 & 0 & 9/8 & 9/7 \\
 11032 & \singleblackcell & 12 & 12 & 0 & 9/8 & 9/7 \\
 37018 & \singlewhitecell & 2 & 02 & 0 & 3/4 & 6/7 \\
 39066 & \singlewhitecell & 1 & 11 & 0 & 3/4 & 6/7 \\
 39394 & \singlewhitecell & 12220 & 12220 & 00 & 21/19 & 21/17 \\
 41114 & \singlewhitecell & 22 & 02 & 0 & 3/4 & 6/7
\end{array}
\]

Three additional morphisms $\varphi$ generate the boundary words of automata with no limiting growth rate.

For rule~$15268$ begun from a single black cell, the boundary word is $\textbf{w}_{15268} = \psi(\varphi^\omega(A))$, where
\begin{align*}
	\varphi &=
	\{
		A \to ABC, \;
		B \to BB, \;
		C \to CC
	\} \\
	\psi &=
	\{
		A \to 220, \;
		B \to 12, \;
		C \to 00
	\}.
\end{align*}
The extremal limit points are $a_\text{inf} = 3/4$ and $a_\text{sup} = 1$.

For rule~$4334$ begun from a single black cell, the morphisms are
\begin{align*}
	\varphi &=
	\{
		A \to AEDBB, \;
		B \to BB, \;
		C \to CC, \;
		D \to DB, \;
		E \to EC
	\} \\
	\psi &=
	\{
		A \to 122, \;
		B \to 22, \;
		C \to 00, \;
		D \to 12, \;
		E \to 10
	\},
\end{align*}
and we have $a_\text{inf} = 6/5$ and $a_\text{sup} = 3/2$.

For rule~$11172$ begun from a single black cell, the morphisms are
\begin{align*}
	\varphi &=
	\{
		A \to AEDB, \;
		B \to BB, \;
		C \to CC, \;
		D \to DB, \;
		E \to EC
	\} \\
	\psi &=
	\{
		A \to 2, \;
		B \to 21, \;
		C \to 00, \;
		D \to 02, \;
		E \to 2\bar{1}
	\},
\end{align*}
and we have $a_\text{inf} = 3/4$ and $a_\text{sup} = 1$.

\section{Automata with irreducible boundaries}\label{irreducible}

Among the $25088$ equivalence classes of $k=2$, $d=4$ cellular automata, $720$ automata evaded all attempts to reduce their boundaries.
Among these $720$, there are only $688$ distinct pictures, since some pairs of inequivalent rules appear to generate the same evolution.
In this section we first comment on the variety of unpredictable behavior found among these boundaries, and then we use tools from Brownian motion to study them more quantitatively.

\subsection{Qualitative taxonomy}

\begin{figure}
	\begin{center}
		\scalebox{.41}{\includegraphics{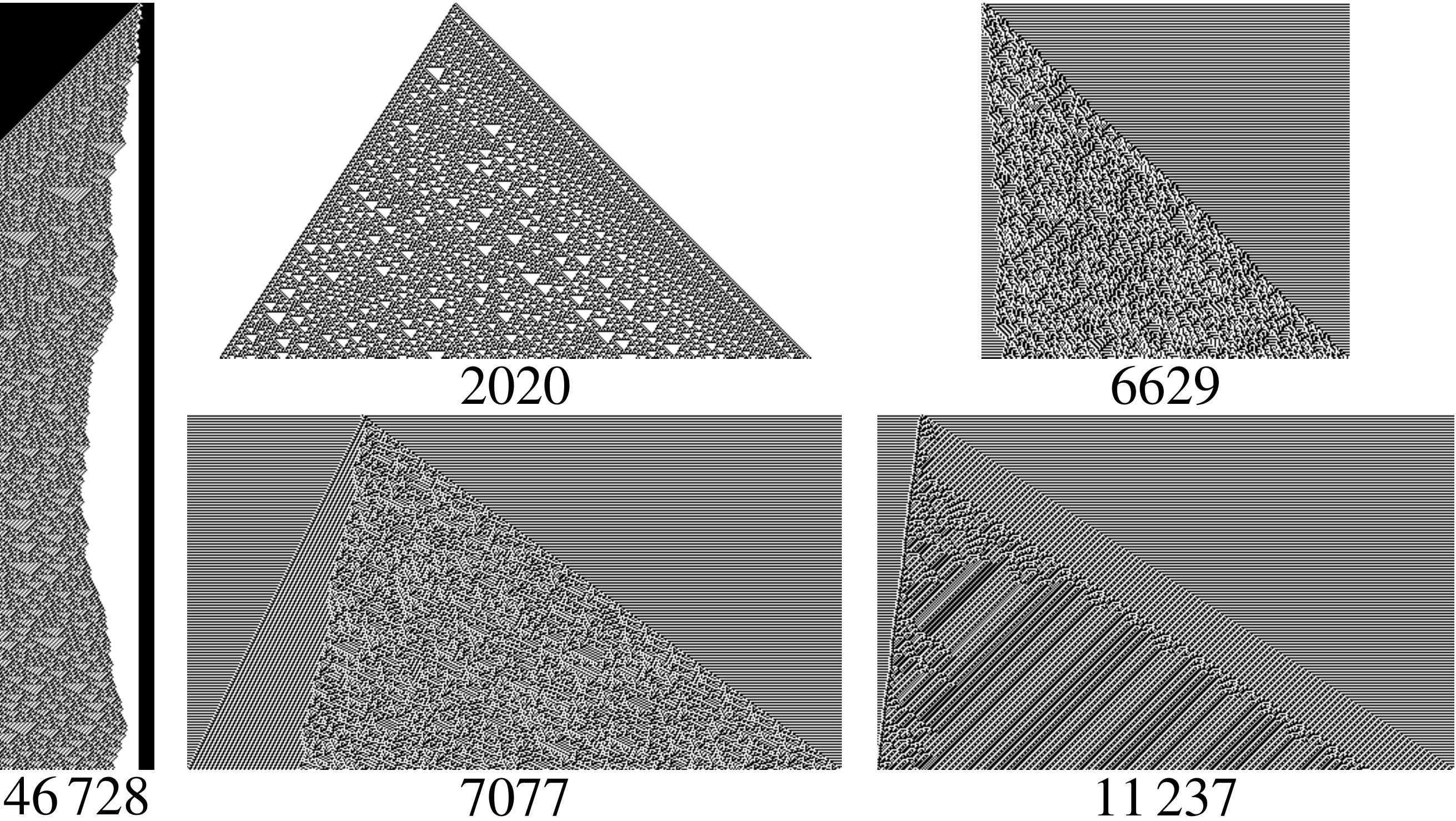}}
	\end{center}
	\caption{
		Examples of qualitatively different kinds of boundaries conjectured to be irreducible. Clockwise from the top: a chaotic interior with a rather stable boundary that is periodic for thousands of time steps at a time ($2020$); rough boundaries on both sides ($6629$); interior particles collide with (and likely prevent the reducibility of) the boundary ($11237$); a light-speed particle outruns a slower, more chaotic boundary ($7077$); an internal boundary resembling a lazy random walk that occasionally hits the (otherwise straight) external boundary ($46728$).
	}
	\label{qualitative_kinds_of_irregular_boundaries}
\end{figure}

Dependence on a fourth neighbor ($d=4$) permits kinds of irregular boundaries that do not occur for the smaller neighborhood $d=3$. Here we attempt to qualitatively survey the different behavior. Some automata, in spite of their chaotic-looking interiors, have stable-looking boundaries, but their chaotic interiors likely prevent the boundaries from stabilizing. The growth of these boundaries may represent an average of the input from the interior. Examples include rules 2020, 2717, 3223, 3493, 5267, 6116, 6773 (begun from one black cell) and 5603 and 5881 (white cell); Figure~\ref{qualitative_kinds_of_irregular_boundaries} shows rule 2020. Some of these boundaries, such as 6773 from black, are periodic for thousands of time steps at a time, but the chaotic interior seems to perpetually break the boundary's reducibility.

Even more exotic and nontrivial boundaries exist. For instance, rules $5673$, $6629$ and $7721$ begun from a single black cell appear to have the rare property of rough boundaries on both sides. Rule 7077 (from black) jettisons a particle at the speed of light to the left, which the slower, apparently chaotic boundary cannot catch. In other cases, interior rather than exterior particles dominate the behavior of the boundaries. For instance, rule 7379 (from either black or white) jettisons diagonal patterns from the left boundary that run nearly parallel to it, while rules 8144 (from black) and 11237 (black or white) create particles that collide with the left boundary at a more oblique angle, which indicates that growth of boundaries may depend on delicate, internal patterns. The nonperiodic internal structures of these automata come remarkably close to the left boundaries; the internal structures seem to persist, causing the boundaries to be nonperiodic.

Most of these boundaries grow with significant average velocity near the speed of light ($d-1$ cells per time step). Others grow as slowly as $0.02$ cells per time step (see Section~\ref{sec:slow}). For instance, rule 46728 from white (shown in Figure~\ref{qualitative_kinds_of_irregular_boundaries}) has an internal boundary resembling a lazy random walk that occasionally collides with the (otherwise straight) external boundary. To quantify descriptions like these, we next study the $688$ unpredictable boundaries by treating them like Brownian motion.

\subsection{Random walk statistics}
To draw an analogy between unpredictable boundaries and random walks, we note that the average growth $\ell(t)/t$ and variance of the difference sequence $\ell(t + 1) - \ell(t)$ of boundaries of cellular automata are analogs of the drift and diffusivity of the Brownian motion of molecular motors~\cite{Kramer 2010}. In light of this parallel, we define the \emph{drift} $U$ to be the average growth rate
\[
U = \lim_{t \rightarrow \infty} \ell(t)/t
\]
and the \emph{diffusivity} $D$ to be the variance of the difference sequence 
\[
D = \lim_{t \rightarrow \infty}\text{Var}(\ell(1) - \ell(0), \ell(2) - \ell(1), \dots, \ell(t + 1) - \ell(t)).
\]
Continuing the analogy with molecular motors, we define a \emph{Pecl\'et number} for boundaries of automata to be the ratio of the drift and diffusivity,
\[
\mathrm{Pe} = \frac{|U|}{D}.
\]
The \emph{Pecl\'et number} $\mathrm{Pe}$ measures the coherence of the boundary~\cite{Kramer 2010}: a large $\mathrm{Pe}$ indicates nearly deterministic movement in a clear direction, whereas a small $\mathrm{Pe}$ indicates a meandering, noisy trajectory. Its inverse $r = 1 / \mathrm{Pe}$ is the \emph{randomness} of the boundary~\cite{Kramer 2010}. 

\begin{figure}
	\begin{center}
		\scalebox{.65}{\includegraphics{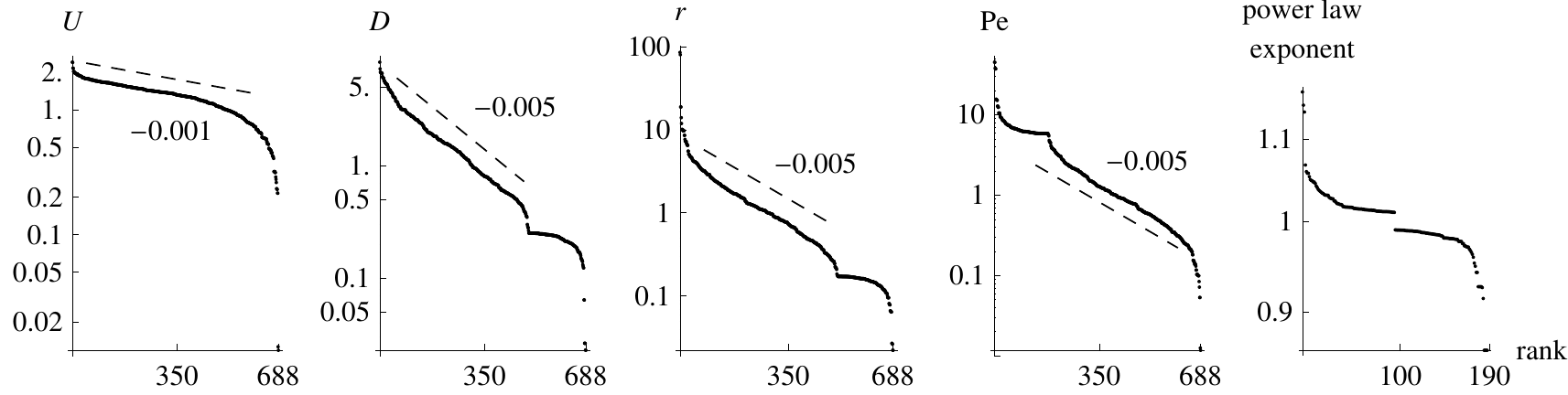}}
	\end{center}
	\caption{
	The four random walk statistics (drift $U$, diffusivity $D$, randomness $r$, Pecl\'et number Pe) of the $688$ irreducible boundaries decay approximately exponentially when sorted in decreasing order. Dashed lines approximate the slopes on log-linear scales. In the rightmost plot, we sort and plot the exponents of the power law fits for the $190$ boundaries deemed to be better fit by a power law ($at^b$ or $at^b+c$) than linear or logarithmic; not shown are the exceptionally small exponents $b= 0.03$ and $0.01$ of rules $7403$ and $7419$.}
	\label{rank-distributions}
\end{figure}

In Figure~\ref{rank-distributions} we plot the distributions of the four random walk statistics ($U,D,r, \mathrm{Pe}$) of the 688 irreducible boundaries. Sorting and plotting these on log-linear scales shows that $U,D,r, \mathrm{Pe}$ decay approximately exponentially over two orders of magnitude among the 688 irregular boundaries. This observation, and others in this section, are robust to changes in the number of time steps $t_\text{max} \in \{500,1500,5000, 10000\}$ of evaluating the automata.\footnote{The values of $U,D$ for almost every automaton change little from calculations up to time $t_\text{max} = 5000$ to calculations up to time $t_\text{max} = 10000$ (e.g., $2/3$ of the diffusivities change by $<0.01$, while $90\%$ change by $<0.05$).} The data stored in \textsc{CellularAutomatonData}~\cite{CellularAutomatonData} is for $t_\text{max}=10^4$, and we show these results throughout this section.

We also fit the boundaries to linear ($at+c$), power law ($a t^b$ and $a t^b+c$), and logarithmic ($a \log(b t)+c$) functional forms. To select the ``best fit'' that maximizes the $R^2$ (for accuracy) and that minimizes the Akaike Information Criterion (AIC) (for parsimony)~\cite{AIC}, we choose the fit that maximizes $R^2 \cdot \exp((\text{AIC}_\text{min}-\text{AIC})/2)$, where $\text{AIC}_\text{min}$ is the minimum AIC among all models~\cite{AIC}.

As expected, for reducible boundaries, the slope $a$ of the linear fit $at+c$ approximately equals both the empirical estimate of the drift, $\ell(t_\text{max})/t_\text{max}$, and the growth rate $a$ in Equation~\ref{periodic boundary word} computed using Equation~\ref{record growth rate}. For boundaries conjectured to be irreducible, the slope $a$ of the linear fit $at+c$ is nearly equal to the empirical estimate of the drift (for $t_\text{max} =10^4$, $a$ and $U$ differ by just $0.002 \pm 0.004$). 

Irrational limiting growth rates are known to exist for cellular automata that compute powers of integers in a certain base~\cite[pages~613--615]{NKS}. However, we did not recognize by visual inspection any irrational numbers among the growth rates of the irregular boundaries, which suggests that they do not exist for $k=2$, $d=4$ rules.
Recognizing exact irrational growth rates is difficult, since one expects $\ell(t_\text{max})/t_\text{max}$ for $t_\text{max} = 10^4$ to agree with the limiting growth rate for at most four or five digits.

No boundaries were deemed best fit by the logarithmic functional form, but $190$ of the $688$ irregular boundaries were deemed best fit by a power law. The exponents $b$ of these power laws all lie in the interval $[0.85,1.17]$, except for the two slowest-growing boundaries, $7403$ and $7419$, both begun from a black cell (with exponents $b= 0.03$ and $0.01$). (For more on the slowest-growing boundaries, see Section~\ref{sec:slow}.) We reject power law fits with exponents $|b-1|< 0.01$, because these are more reasonably deemed linear fits. We conclude that nearly all the boundaries that grow as power laws have exponents near 1. Exponents above 1 occur when the parameter $a<1$, which cannot be accurate for sufficiently large $t$ because $\ell(t) \le 3t + 1$ for all $d=4$ automata begun from a single-cell initial condition. Neither adjusting $t_\text{max}$ nor dropping tens or hundreds of the first boundary lengths (to allow for a transient) eliminates the power law exponents larger than $1$. This illustrates the difficulties of fitting irregular boundaries to functional forms using standard nonlinear fitting algorithms. 

The drift $U$ and diffusivity $D$ characterize what kinds of random walks these irregular boundaries behave like. Notably, one quarter of the 688 automata have diffusivity $0.15 < D < 0.25$, which creates a ``knee'' in Figure~\ref{rank-distributions}. For comparison, a simple random walk with steps $1,-1$ occurring with probability $p, 1-p$ has variance 0.25 for $p=\frac{1}{4} \left(2-\sqrt{3}\right) \approx 0.067$. Such a random walk moves rather coherently in a certain direction, reflected by its large Pecl\'et number $\mathrm{Pe} = 2 \sqrt{3} \approx 3.5$ that is also common among the irregular boundaries.

\begin{figure}
	\begin{center}
		\scalebox{.5}{\includegraphics{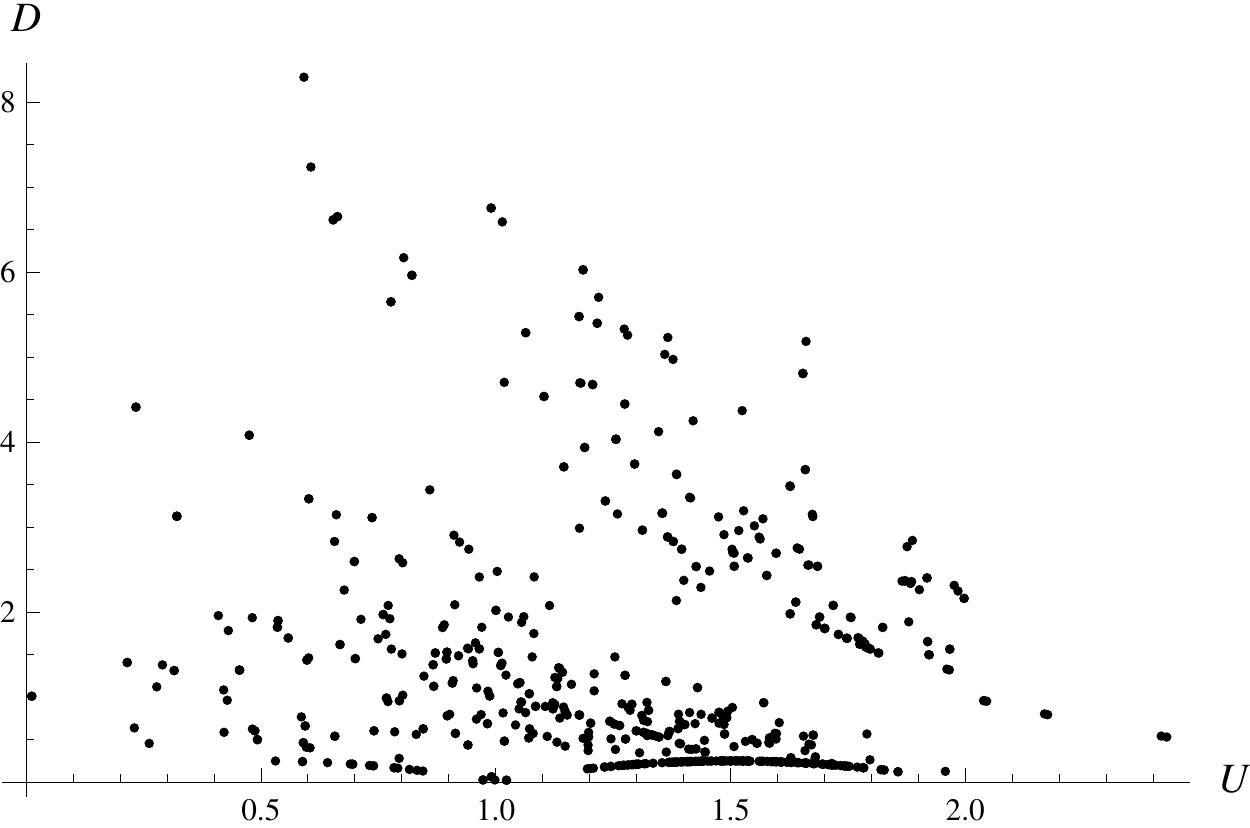}}	
	\end{center}
	\caption{An unexpected gap in the relationship between diffusivity $D$ and drift $U$ (computed for $10^4$ steps) suggests a threshold exists in the behavior of irreducible boundaries of cellular automata: they either grow erratically and quickly or more deterministically and slowly.}
	\label{DvsU}
\end{figure}

Turning our attention to the drift $U$ and diffusivity $D$ of all $688$ irregular boundaries, we find a gap in the scatter plot of $D$ and $U$ in Figure~\ref{DvsU}. This gap suggests the existence of a threshold: irreducible boundaries of automata either grow quickly and erratically (large $U,D$; upper-right region of Figure~\ref{DvsU}) or more slowly and deterministically (small $U,D$; lower-left region of Figure~\ref{DvsU}). This scatter plot and its gap do not change qualitatively for different numbers of time steps $t_\text{max}$.

\subsection{Slow growth}\label{sec:slow} Fast boundary growth is common: the mean growth rate among the boundaries conjectured to be irreducible is large, $\langle U \rangle \approx 1.27$. Slow growth, by contrast, is delicate and rare (see the sparse region $U<0.5$ in Figure~\ref{DvsU}). Table~\ref{slowest-rules} shows the ten automata that grow most slowly among the 688 automata with apparently irreducible boundaries. The last column depicts the initial terms of the sequences $\ell(t)$.

\begin{table}[htdp]
\begin{center}
\begin{tabular}{c c c c}
rule & initial condition & drift $U$ & $\ell(t)$ for $t \leq 500$ \\ \hline
 7403 & $\singleblackcell$ &  0.021404 & \includegraphics[scale=1]{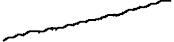}\\
 7419 & $\singleblackcell$ & 0.023805 & \includegraphics[scale=1]{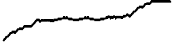}\\
 2295 & $\singleblackcell$ & 0.210042  & \includegraphics[scale=1]{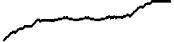}\\
 2295 & $\singlewhitecell$ & 0.210042  & \includegraphics[scale=1]{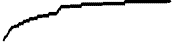}\\
 11411 & $\singlewhitecell$ & 0.230046  & \includegraphics[scale=1]{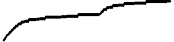}\\
 11411 & $\singleblackcell$ & 0.230446  & \includegraphics[scale=1]{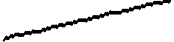}\\
 38538 & $\singlewhitecell$ & 0.233647  & \includegraphics[scale=1]{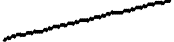}\\
 34490 & $\singleblackcell$ & 0.264053  & \includegraphics[scale=1]{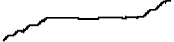}\\
 34458 & $\singleblackcell$ & 0.266853  & \includegraphics[scale=1]{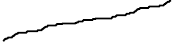}\\
 1690 & $\singleblackcell$ & 0.296859  & \includegraphics[scale=1]{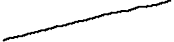}\\
 \end{tabular}
\end{center}
\caption{The ten slowest of the presumably irreducible boundaries (in the first $5000$ time steps).}
\label{slowest-rules}
\end{table}

\begin{figure}
	\begin{center}
		\includegraphics{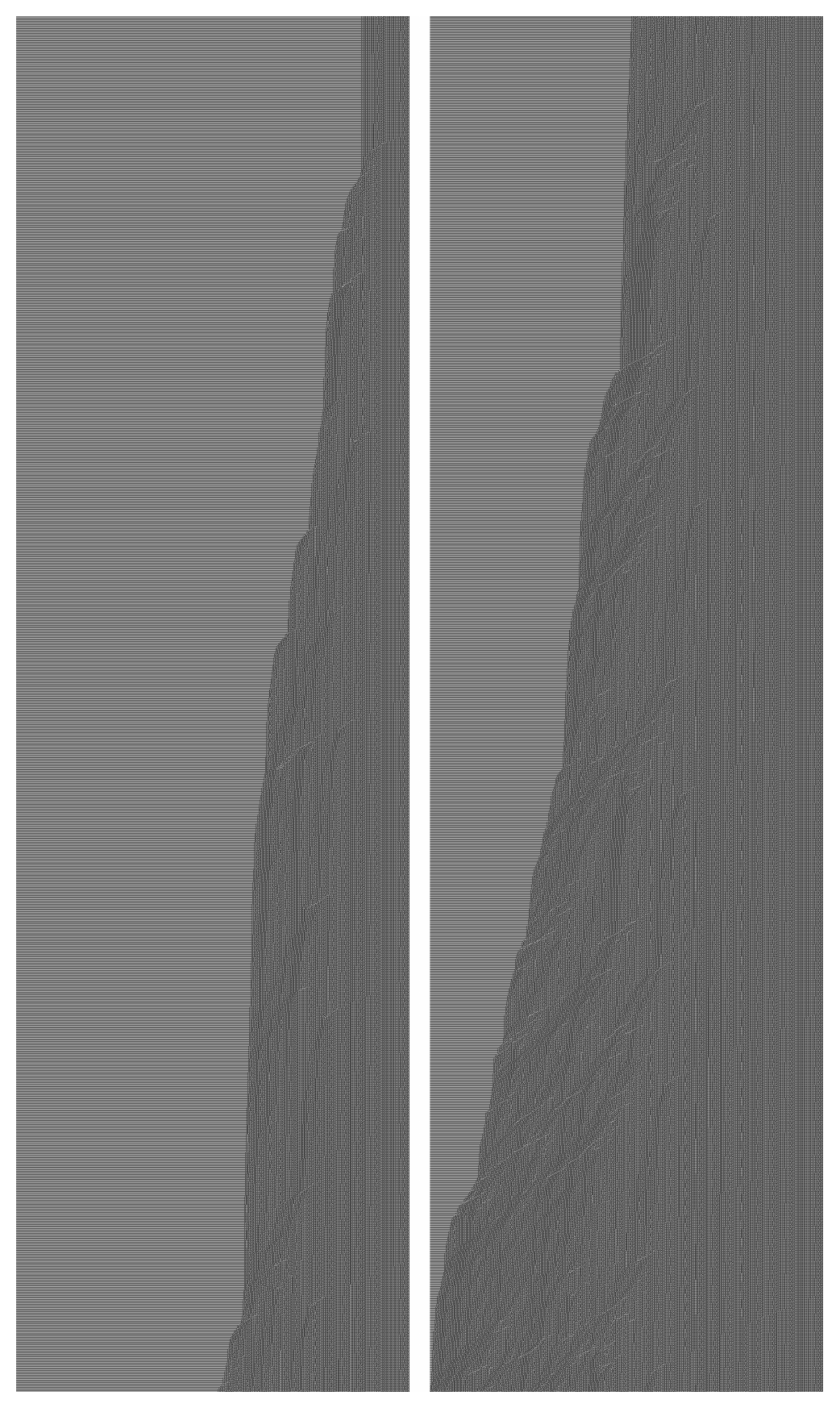}
	\end{center}
	\caption{Steps $524000$ through $534001$ of rule $7403$, broken up into two columns.  After growing to just $174$ cells wide in the first $524000$ steps, the automaton begins to grow much more rapidly at $t=524557$, reaching length $1429$ at time $t=534001$.}
	\label{rule_7403}
\end{figure}

The very slowest automaton (at least in the first $5000$ steps) is rule~$7403$ begun from $\singleblackcell$, shown in Figure~\ref{rules_7403,7419}, which does something quite surprising. Its boundary continues to grow slowly for more than half a million steps, reaching only $\ell(524557) = 174$. After step $524557$ the growth increases dramatically, reaching length $277$ at step $525000$ and length $36819$ at step $10^6$. So while the average growth rate for the range $0 \leq t \leq 500000$ is $0.000348$, the average growth rate for $500000 \leq t \leq 10^6$ is $0.073290$, as if for some reason a growth rate as low as $0.000348$ is not sustainable.
Figure~\ref{rules_7403,7419} (bottom) and Figure~\ref{rule_7403} show the point at which the growth rate changes.
We have no explanation for this behavior.

\begin{figure}
	\begin{center}
		\includegraphics{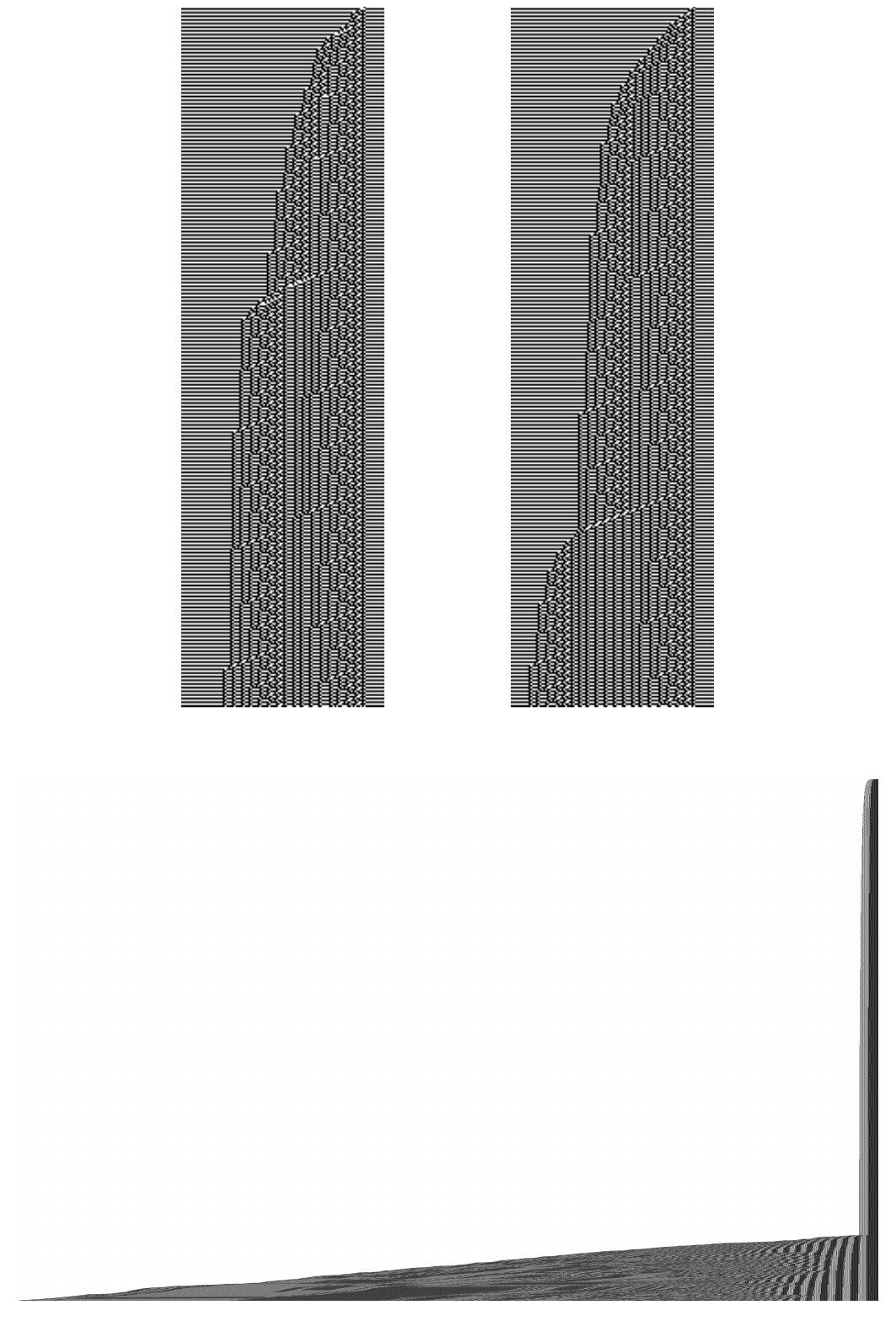}
	\end{center}
	\caption{
		Top: Rules $7403$ and $7419$ begun from $\singleblackcell$, the slowest-growing $k = 2$, $d = 4$ automata with single-cell initial conditions.
		Bottom:  The first $6\times10^5$ steps of rule $7403$, sampled every $128$ steps, illustrate the explosion of boundary growth at step $524557$.
	}
	\label{rules_7403,7419}
\end{figure}

The boundary of rule~$7419$ begun from $\singleblackcell$ (also shown in Figure~\ref{rules_7403,7419}) also exhibits extended slow growth. Unlike rule~$7403$, however, its growth rate does not appear to suddenly increase. Due to its relatively short rows, one can quickly evolve it for a large number of steps. For example, we compute $\ell(10^8) = 271$, and one suspects that the growth of this automaton is not linear in general but is better modeled by $a t^b$. We have no explanation for this continued slow growth either.

We remark that the pictures generated by rules~$7403$ and $7419$, shown in Figure~\ref{rules_7403,7419}, resemble each other significantly. They largely consist of vertical lines, with structure reminiscent of counting in binary. Further work should be undertaken to understand these rules and to determine the extent to which they are reducible.

\subsection{Potential for universality}

Rule~$7555$ is interesting as a potentially programmable rule and hence a candidate for universality.
Begun from either initial condition, the picture that rule~$7555$ generates strongly suggests that it performs some kind of arithmetic, with clear particles of varying slopes at times passing through each other and at other times interacting. Its left boundary depends sensitively on the computations being performed in the interior, and, for example, after changing position thirteen times in the range $10000 \leq t \leq 20000$ when begun from $\singlewhitecell$, it remains constant for more than $3000$ steps beginning at step $20555$.

\section{An automaton with no growth exponent}\label{no exponent}

In this section we show that it is not possible in general to assign a growth function $t^b$ to a cellular automaton. In particular, we construct an automaton such that the limiting growth exponent
\[
	\lim_{t \to \infty} \log_t \ell(t)
\]
does not exist.

The idea is to take rule~$106$ begun from $\twoblackcells$ (shown in Figure~\ref{rules_45,107,106}), which grows like $\sqrt t$, and to graft onto it an automaton that roughly squares the length of a row. We set up the squaring rule to be activated at certain steps, causing the sequence $\ell(t)$ to grow to be on the order of $t$, and then we allow it to fall back to the boundary of rule~$106$ on the order of $\sqrt t$ before being activated again. As a result, the sequence $\ell(t)$ oscillates between square-root growth and linear growth and satisfies
\[
	\liminf_{t \to \infty} \log_t \ell(t) = \frac{1}{2},
	\qquad
	\limsup_{t \to \infty} \log_t \ell(t) = 1.
\]

\begin{figure}
	\begin{center}
		\includegraphics{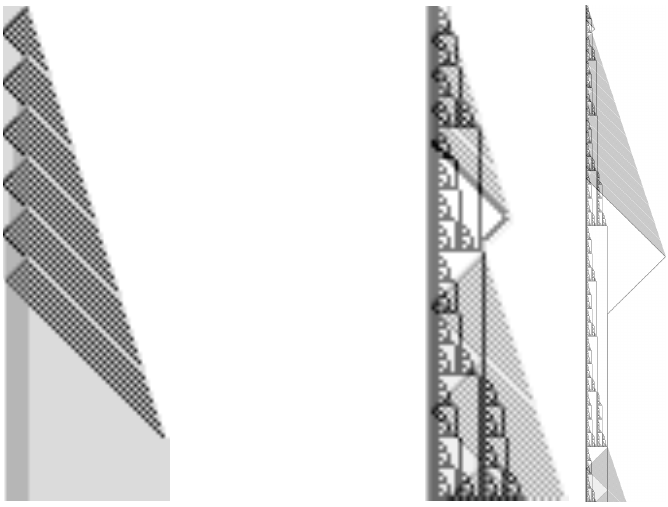}
	\end{center}
	\caption{
		Left: A cellular automaton that squares integers, shown here squaring $\ell - 1 = 6$.
		Right: A cellular automaton with no growth exponent, shown for $128$ steps and $1152$ steps.
	}
	\label{squaring_and_composite_automata}
\end{figure}

A squaring rule that works by repeated addition was given by Wolfram~\cite[page~639]{NKS} using $k = 8$ and $d = 3$. Begun from the initial condition
\[
	\cdots000\s\underbrace{11\cdots11}_{\ell-1}\s3000\cdots,
\]
this rule produces a row of length $\ell^2 - \ell$ after $3 \ell^2 - 5 \ell$ steps.
Figure~\ref{squaring_and_composite_automata} shows the automaton squaring the integer $6$.

A $k_1$-color rule and a $k_2$-color rule can be combined into a single $(k_1 k_2)$-color rule that can be thought of as their direct product and that can run the two rules in parallel. Since of course we do not want the two rules to run completely independently, we modify the composite rule so that there is some interaction.
In particular, modifications to the squaring automaton, including the addition of one color, inhibit future squarings until the current squaring is finished and the automaton has shrunk to the width of rule~$106$.
Hence our composite rule uses $2 \times 9 = 18$ colors. The broad outline is as follows.

Step~$3$ in rule~$106$ consists of four black cells.
We choose the initial condition so that the squaring automaton is first activated on step~$4$.
Since the squaring needs to be activated locally, we modify the squaring automaton so that it squares a row using only information from its two endpoints rather than from the entire interval of cells.
The relevant interval for squaring on step~$4$ has length $\ell = 5$, so the squaring automaton takes $3 \ell^2 - 5 \ell = 50$ steps to square.
From the time the squaring begins until it finishes, the squaring automaton runs independently of rule~$106$.

After the squaring completes at step~$54$, we must clear the cells used by the squaring automaton. To do this, we add a new color to mark the leftmost nonempty column. When the last addition is complete, we send out a particle from this column that travels to the right and clears the cells involved in squaring.

When the clearing particle reaches the rightmost remnant of the squaring automaton, we trigger a particle traveling back to the left to signify that the next squaring can begin. When this particle first encounters a structure from rule~$106$, it stops propagating to the left and remains in that column to trigger the next squaring when rule~$106$ next has two adjacent black cells at the right endpoint, and the process begins again.

The result is a rule with $k = 18$ and $d = 4$, begun from the initial condition $\cdots0002899003000\cdots$.
Figure~\ref{squaring_and_composite_automata} shows two images of this automaton. The complete rule instructions are available in \textsc{CellularAutomatonData}~\cite{CellularAutomatonData}. Even though both rule~$106$ and the squaring automaton are functions of $3$ cells, it is necessary to shear one of the rules relative to the other to bring their structures into alignment, hence $d = 4$.

We now verify by induction that triggering the initial squaring on step~$3$ enables one to easily determine when all future squarings will occur. (For other initial triggering steps this is not the case.) We claim that squarings are triggered precisely on steps $2^{4 \alpha + 2} - 1$ for $\alpha \geq 0$.

For $\alpha = 0$ the squaring at step $3$ is guaranteed by our choice of initial condition.

Inductively, assume that a squaring is triggered on step $2^{4 \alpha + 2} - 1$ for a fixed $\alpha$.
On step $2^{4 \alpha + 2} - 1$, rule~$106$ has a solid black row of length $3 \cdot 2^{2 \alpha} + 1$.
The squaring rule begins squaring $3 \cdot 2^{2 \alpha} + 2$ on the following step and reaches maximum length $9 \cdot 2^{4 \alpha} + 9 \cdot 2^{2 \alpha} + 5$ on step $31 \cdot 2^{4 \alpha} + 21 \cdot 2^{2 \alpha} + 2$. The length is maximal for three steps, and then the length shrinks one cell per step until the particle reaches the boundary of rule~$106$; this occurs at step $5 \cdot 2^{4 \alpha + 3} + 9 \cdot 2^{2 \alpha + 1} + 3$, because the length of rule~$106$ is $3 \cdot 2^{2 (\alpha + 1)} + 1$ for all $2^{4 \alpha + 5} \leq t \leq 2^{4 \alpha + 6} - 1$, and one checks that $2^{4 \alpha + 5} < 5 \cdot 2^{4 \alpha + 3} + 9 \cdot 2^{2 \alpha + 1} + 3 < 2^{4 \alpha + 6} - 2$. For $2^{4 \alpha + 5} \leq t \leq 2^{4 \alpha + 6} - 2$, the right endpoint of rule~$106$ is a single black cell, and the next occurrence of two adjacent black cells is on step $2^{4 (\alpha + 1) + 2} - 1$.

It follows that the subsequence of the steps where squarings begin has limiting exponent
\[
	\lim_{\alpha \to \infty} \frac{\log (3 \cdot 2^{2 \alpha} + 1)}{\log (2^{4 \alpha + 2} - 1)} = \frac{1}{2},
\]
and the subsequence of the steps where squarings end has limiting exponent
\[
	\lim_{\alpha \to \infty} \frac{\log (9 \cdot 2^{4 \alpha} + 9 \cdot 2^{2 \alpha} + 5)}{\log (31 \cdot 2^{4 \alpha} + 21 \cdot 2^{2 \alpha} + 2)} = 1.
\]

\section{Conclusions and open questions}\label{conclusion}

In this paper we have inventoried the boundaries of all cellular automaton rules using $k = 2$ colors and depending on $d = 4$ cells when begun from a single cell on a constant background.
Within this rule space we have encountered several kinds of behavior not seen in smaller spaces. In particular, we find fractal boundaries described by morphic words. By studying the unpredictable boundaries as if they were random walks, we find approximately exponential distributions of the mean and variance of the boundaries' growth and a possible separation into two classes of automata, ones that grow quickly and erratically and others that grow slowly and more deterministically. 

For simplicity, we have restricted our attention in many ways. We have only considered the two initial conditions $\singleblackcell$ and $\singlewhitecell$. A more general study of $k = 2$, $d = 4$ rules will consider other initial conditions and attempt to determine to what extent each rule has a representative growth rate.
More generally still, one can consider initial conditions with backgrounds that are not constant but are periodic, because there still exists a natural notion of the length of a row. Finally, the rule space we studied is big, but it is not huge, and one can imagine performing similar analyses on larger spaces of cellular automata. We hope that researchers in fact do all of these things, and we have designed \texttt{CellularAutomatonData} to scale to these more general settings.

Another topic to be addressed is the issue of distinct automata that nevertheless generate the same evolution (or an evolution equivalent under reflection or permutation of colors) because certain local configurations of cells do not appear. For example, rules~$34394$ and $39780$ can generate identical evolutions, as mentioned in Section~\ref{square-root growth}. At the beginning of Section~\ref{irreducible} we encountered this phenomenon again. (Although we did not mention it earlier, among the $688$ distinct evolutions generated by the $720$ irreducible automata there appear to be only $658$ distinct boundary words.) The prevalence of equivalent evolutions generated by inequivalent rules suggests that one should use more complex initial conditions to distinguish such rules. One possible criterion for a representative initial condition for a given rule is that all $k^d$ local configurations that can (for some initial condition) occur infinitely often in an evolution do occur infinitely often.

This paper concerns external boundaries, which are simply special cases of general boundaries between distinct regions of a cellular automaton's evolution. The advantage of studying external boundaries is the ease of defining and therefore programmatically detecting them. However, internal boundaries (or particles) are common in automata, and several information-theoretic measures have been used to detect them~\cite{Wuensche 1999,Lizier 2008,Helvik 2004,Shalizi 2006} and their collisions~\cite{Lizier 2010}. We expect that our automated and manual methods could inform a study of general boundaries. Conversely, information-theoretic tools for internal boundaries may be applied to external boundaries to systematically measure, for instance, how much they store and process information~\cite{Hanson 1997,Hordijk 2001}.

In most cases, the cellular automaton data we computed is empirical and has not been formally proved to be correct. (We welcome any corrections.) Of course ideally we would like to have proofs. The automata with morphic boundary words that are not eventually periodic are few enough, at least in the space $k=2$, $d=4$, that it is reasonable to attempt to prove manually that each boundary word is described by the morphism claimed. On the other hand, for the $24287$ automata with eventually periodic boundary words, obtaining proofs by hand is not reasonable, and one must develop automated techniques for examining a rule and initial condition to determine (rigorously) the growth rate and the eventual period length. Of course, the question of whether the boundary word is eventually periodic is likely undecidable in general. However, a symbolic approach capable of proving a large number of growth rates would be of great interest.

From the results in this paper, several natural questions arise regarding the growth of cellular automata.
\begin{itemize}

\item
Which morphic words occur as the boundary word of a cellular automaton?

\item
For what real numbers $0 \leq b \leq 1$ is there a cellular automaton with limiting growth exponent $b = \lim_{t \to \infty} \log_t \ell(t)$?
Schaeffer~\cite{Schaeffer} has recently constructed cellular automata with row lengths that grow like $t^{1/m}$ for any integer $m \geq 3$, and $t^{\log_2 \phi}$ where $\phi = (1+\sqrt{5})/2$.

\item
Schaeffer~\cite{Schaeffer} has also constructed an automaton with $\ell(t) = O(\sqrt t \log t)$.
What can be said in general about possible and impossible growth functions?

\item
How does the growth of an automaton depend on $k$ and $d$?
For example, what is the smallest nonzero rational growth rate that occurs for given $k$ and $d$?

\end{itemize}
These and other questions indicate the breadth of mathematics and experimentation to be done on the boundaries of cellular automata.

\section*{Acknowledgments}

We thank Hector Zenil for contributions at the NKS Summer School 2009 and Janko Gravner for useful discussion. The first author was supported by the Defense Threat Reduction Agency, Basic Research Award HDTRA1-10-1-0088, and by the Department of Defense (DoD) through the National Defense Science \& Engineering Graduate Fellowship (NDSEG) Program.
The second author was supported in part by NSF grant DMS-0239996.

\end{document}